\documentclass[a4paper,11pt]{article}
\usepackage{jheppub}
\usepackage{float}

\begin{document}
\title{Clarifying Kasner dynamics inside anisotropic black hole with vector hair}

\author{Rong-Gen Cai$^{1,2,3,4}$, Mei-Ning Duan$^{2,3}$, Li Li$^{1,2,3}$, Fu-Guo Yang$^{1,2,5}$}

\affiliation{$^{1}$School of Fundamental Physics and Mathematical Sciences, Hangzhou Institute for Advanced Study, UCAS, Hangzhou 310024, China.}

\affiliation{$^{2}$School of Physical Sciences, University of Chinese Academy of Sciences, No.19A Yuquan Road, Beijing 100049, China.}

\affiliation{$^{3}$CAS Key Laboratory of Theoretical Physics, Institute of Theoretical Physics, Chinese Academy of Sciences, P.O. Box 2735, Beijing 100190, China.}

\affiliation{$^{4}$School of Physical Science and Technology, Ningbo University, Ningbo, 315211, China}

\affiliation{$^{5}$International Centre for Theoretical Physics Asia-Pacific, University of Chinese Academy of Sciences, 100190 Beijing, China}

\emailAdd{cairg@itp.ac.cn}
\emailAdd{duanmeining@itp.ac.cn}
\emailAdd{liliphy@itp.ac.cn}
\emailAdd{yangfuguo@ucas.ac.cn}

%\pacs{PACS}
%\keywords{keywords}
%\preprint{preprint}
%%%%%%%%%%%%%%%%%%%%%%%%%%%%%%%%%%%%%%
%\begin{abstract}
%%%%%%%%%%%%%%%%%%%%%%%%%%%%%%%%%%%%%%
\abstract{We investigate the internal dynamics of hairy black holes in the Einstein-Maxwell-vector theory. The development of the charged vector hair necessarily removes the inner Cauchy horizon and results in an anisotropic black hole ending at a spacelike singularity. The far interior evolution is characterized by the chaotic alternations  among different Kasner epochs. We show that those late interior time dynamics of bounces and epochs can be captured analytically. In particular, we obtain three distinct types of transformation laws for the alternation of Kasner epochs analytically, including the Kasner inversion, Kasner transition, and Kasner reflection. Our analytical approach is corroborated by numerical solutions to the full equations of motion. Moreover, we provide a clear explanation of the alternation of Kasner epochs found numerically in the literature.}

%\end{abstract}
%%%%%%%%%%%%%%%%%%%%%%%%%%%%%%%%%%%%%%
\maketitle
%\tableofcontents
\flushbottom

%%%%%%%%%%%%%%%%%%%%%%%%%%%%%%%%%%%%%%
\noindent
\newpage

\section{Introduction}\label{Sec:Intro}
Since the concept of black holes was proposed, significant progress has been made in understanding their internal structure. A landmark study is the singularity theorem by Penrose and Hawking~\cite{Penrose:1964wq,Hawking:1970zqf}, which guarantees that some sort of geodesic incompleteness (singularity) occurs inside any black hole whenever reasonable energy conditions are satisfied, irrespective of the spacetime symmetry. Thus, black hole formation is a robust prediction of the general theory of relativity. However, this theorem does not provide any detailed description of the dynamics near the singularity. Almost at the same time, Belinskii, Khalatnikov and Lifshitz (BKL) solved Einstein's equations near a spacelike singularity and found that the generic solution is a Kasner spacetime with chaotic oscillations~\cite{Lifshitz:1963ps,Belinsky:1970ew}. It was then shown that the ergodic dynamics of the far interior emerges together with a time-independent Hamiltonian using ``cosmological billiards" in which the evolution of the metric components is mapped onto the motion of a particle on the hyperbolic plane, with the alternation of the Kasner exponent caused by billiard bouncing~\cite{Damour:2002et}.
Nevertheless, this method becomes not applicable in the case for which interactions of matter content become important when writing the Hamiltonian for the particle motion, see \emph{e.g.} the scalar field with super-exponential or even catastrophic potentials~\cite{Cai:2023igv,Hartnoll:2022rdv}. Interestingly, some universal restrictions on the number of horizons were found more recently~\cite{Yang:2021civ} and in some cases the no inner-horizon theorems can be proved for hairy black holes~\cite{Cai:2020wrp,An:2021plu,Cai:2021obq}.

In recent years, exploring the internal structure of a black hole has garnered great interest stimulated by the holographic duality, although the holographic meaning of the black hole interior remains unclear up to now, see earlier work~\cite{Kraus:2002iv,Fidkowski:2003nf,Festuccia:2005pi}. Begin with the Schwarzschild-AdS interior, many studies considered homogeneous, relevant scalar sources at the AdS boundary, changing the Kasner exponents towards the interior singularity~\cite{Frenkel:2020ysx,Hartnoll:2020rwq,Hartnoll:2020fhc}. Some interesting phenomena were observed in the interior including collapse of the Einstein-Rosen bridge, Josephson oscillations and Kasner geometries. At late interior times, the emergence of Kasner scaling towards the singularity is a common feature. Depending on the interactions, a further phenomenon appearing is bounces between different Kasner epochs~\cite{Hartnoll:2020rwq,Hartnoll:2020fhc,Wang:2020nkd,Dias:2021afz,Caceres:2022smh,Sword:2021pfm}.
There are scenarios yielding a never-ending chaotic alternation of Kasner epochs towards the singularity~\cite{An:2022lvo,Hartnoll:2022rdv,Cai:2023igv}. More recently, the classification of the interior dynamics of those hairy black holes was given in~\cite{Cai:2023igv}, where analytic expressions for the Kasner inversion and Kasner transition were obtained, and more complicated behaviors beyond Kasner epochs were uncovered.

The interior dynamics for the black holes with vector hair were considered, which is more challenging than the scalar case due to the spatial anisotropy. Before reaching the singularity, several intermediate regimes develop, including the Einstein-Rosen bridge contracting towards the singularity, the oscillations of vector condensate and the anisotropy of spatial geometry, and a sequence of Kasner epochs connected by bounces. The mechanism for the former two behaviors were clearly explained~\cite{Cai:2021obq}. Nevertheless, the alternation of Kasner epochs was only observed numerically~\cite{Cai:2021obq,Sword:2022oyg}. Although it was described geometrically as collisions between p-form ``walls''~\cite{Henneaux:2022ijt} (see also~\cite{DeClerck:2023fax}), a comprehensive understanding of the dynamics has not yet to be established. In particular, the empirical relations for the alternation of Kasner epochs were reported in~\cite{Cai:2021obq} summarized by a large number of numerical examples. Near some specific temperature values, there develop numerous oscillations extending deep towards the black hole singularity with the amplitude of oscillations increasing~\cite{Sword:2022oyg}. The interior could yield further interesting behaviors upon exploring the parameter space. This paper aims to provide insights into the internal structure of anisotropic black holes with vector hair and to give a clear interpretation on the numerical results reported in~\cite{Cai:2021obq,Sword:2022oyg}. More precisely, we will consider the charged vector case known as holographic p-wave superconductor~\cite{Cai:2013pda,Cai:2013aca}. Based on the method developed in our previous work~\cite{Cai:2023igv}, we will classify three different types of Kasner alternations analytically, including the Kasner transition, Kasner inversion and Kasner reflection in general spacetime dimensions. This provides a different way from the cosmological billiard  to understand the black hole interior dynamics. In $(3+1)$ dimensional spacetime, the corresponding transformation rules for the Kasner exponent agree with the known formula obtained via the cosmological billiard, confirming the correctness of our analysis. We will present numerical checks by solving the full equations of motion. We manage to provide a clear explanation of the alternation of Kasner epochs found numerically in the literature~\cite{Cai:2021obq,Sword:2022oyg}.

The rest of our work is organized as follows. In Section~\ref{Sec:ModelAndInterior}, we introduce the gravitational model and derive the equations of motion. By constructing a conserved charge, we show that vector hair eliminates the Cauchy horizon inside such hairy black holes. In Section~\ref{SubSec:KasStructure}, we analytically derive the laws of Kasner transformation and verify the self-consistency of the analytic approximation together with explicit numerical checking. Based on our findings, in Section~\ref{Sec:old}, we elucidate the interior dynamics observed numerically in the literature. We conclude with some discussion in Section~\ref{Sec:DissAndCon}.

\section{Setup}\label{Sec:ModelAndInterior}
\subsection{Model and Equations of Motion}\label{SubSec:GenSetup}
We begin with a (d+1)-dimensional Einstein-
Maxwell-vector theory that involves a complex vector field $\rho_\mu$ charged under a U(1) gauge field $A_\mu$. The action reads~\cite{Cai:2013pda,Cai:2013aca}
\begin{equation}\label{VectorModel}
S = \int \mathrm{d}^{d+1}x \sqrt{- g} \left( R -2 \Lambda +\mathcal{L}_m\right)\,,\quad \mathcal{L}_m=-\frac{1}{4}F_{\mu\nu}F^{\mu\nu}-\frac{1}{2}\rho_{\mu\nu}^{\dagger}\rho^{\mu\nu}\,,
\end{equation}
where $R$ is the Ricci scalar, and $\Lambda = -\frac{d(d-1)}{2L^{2}}$ is the cosmological constant with $L$ the radius of curvature of AdS. $\rho_{\mu\nu}$ is defined by $\rho_{\mu\nu}=2D_{[\mu}\rho_{\nu]}$ with the covariant derivative $D_{\mu}=\nabla_{\mu}-iqA_{\mu}$. The Maxwell field strength is $F_{\mu\nu}=2\nabla_{[\mu}A_{\nu]}$. We have dropped the mass term since the Kasner alternations inside black holes with charged vector hair were numerically studied for massless case~\cite{Cai:2021obq,Sword:2022oyg}.~\footnote{The contribution from the mass term in the far interior dynamics could be neglected in most cases. One can also include the magnetic moment term $iq \gamma\rho_{\mu}^{\dagger}\rho^{\nu}F^{\mu\nu}$ that plays an important role in the case with an applied magnetic field~\cite{Cai:2013pda}. In the present work this term plays no role in the interior dynamics.}

The ansatz for the matter content and metric is set as follows:
\begin{equation}\label{GenAnsatz}
\begin{aligned}
\mathrm{d}s^2 = \frac{1}{z^2}\left(-f(z) e^{-\chi(z)} \mathrm{d}t^2 + \frac{1}{f(z)} \mathrm{d}z^2 + e^{2(d-2)\zeta(z)} \mathrm{d}x^2 + \frac{1}{e^{2\zeta(z)}} \mathrm{d}\Sigma^2_{d-2} \right)\,,\\
A_{\nu}\mathrm{d}x^{\nu}=A_{t}(z) \mathrm{d}t\,,\quad \rho_{\mu}\mathrm{d}x^{\mu}=\rho_{x}(z)\mathrm{d}x\,,
\end{aligned}
\end{equation}
where $\mathrm{d}\Sigma^2_{d-2}$ denotes the line element for $(d-2)$-dimensional Euclidean space.
Note that the geometry is anisotropic due to the formation of a vector condensation along the $x$-direction. In our coordinate system, the AdS boundary locates at $z=0$ and the singularity would be $z\rightarrow\infty$. At the event horizon of the hairy black hole $z=z_H$, the blackening function $f(z)$ vanishes. Substituting the above ansatz into the action results in the following equations of motion.
\begin{equation}\label{EoM:zeta}
\zeta''=-\left(\frac{1}{z}+\frac{h'}{h}\right)\zeta'-\frac{1}{d-1}(z\rho_{x}')^2\mathrm{e}^{-2(d-2)\zeta}+\frac{q^2A_{t}^2\rho_{x}^2}{(d-1)z^{2d-2}h^2}\mathrm{e}^{-2(d-2)\zeta}\,,
\end{equation}
\begin{equation}\label{EoM:rho}
(z\rho_{x}')'=2\left((d-2)\zeta'-\frac{1}{z}-\frac{h'}{2h}\right)z\rho_{x}'-\frac{q^2A_{t}^2\rho_{x}}{z^{2d-1}h^2}\,,
\end{equation}
\begin{equation}\label{EoM:At}
\left(\frac{\mathrm{e}^{\chi/2}A_{t}'}{z^{d-3}}\right)'=\frac{2q^2A_{t}\rho_{x}^2}{z^{2d-3}h}\mathrm{e}^{-2(d-2)\zeta}\,,
\end{equation}
\begin{equation}\label{EoM:chi}
\chi'=2(d-2)z\zeta'^2+\frac{2}{d-1}z(z\rho_{x}')^2\mathrm{e}^{-2(d-2)\zeta}+\frac{2q^2A_{t}^2\rho_{x}^2}{(d-1)z^{2d-3}h^2}\mathrm{e}^{-2(d-2)\zeta}\,,
\end{equation}
\begin{equation}\label{EoM:h}
h'=\frac{\mathrm{e}^{-\chi/2}}{d-1}\left(\frac{2\Lambda}{z^{d+1}}+\frac{\mathrm{e}^{\chi}A_{t}'^2}{2z^{d-3}}\right)\,,
\end{equation}
where the prime denotes the derivative with respect to the radial coordinate $z$ and  $h = z^{-d}\mathrm{e}^{-\chi/2}f$ is introduced for later convenience. 

Solving the equations~\eqref{EoM:zeta}-\eqref{EoM:h} analytically is impossible, but we can obtain approximate solutions to gain insight into the internal structure of vectorized hairy black holes. In Section~\ref{SubSec:KasStructure}, we will find a self-consistent approximate solution to the above equations of motion.

\subsection{Proof of No Inner Horizon}\label{SubSec:Noether Charge}
Following~\cite{Cai:2021obq}, we derive the radially conserved quantity by exploiting the scaling symmetry of the system. Substituting the ansatz~\eqref{GenAnsatz} into the action~\eqref{VectorModel}, we obtain the following effective action:
\begin{equation}\label{EffectAction}
\begin{aligned}
S_{eff} =& \frac{\mathcal{V}_{d}}{\kappa_{(d+1)}^{2}}\int dz \mathcal{L}_{eff}(f,f',f'';\chi,\chi',\chi'';\zeta,\zeta';\rho_{x},\rho_{x}';A_{t},A_{t}';z)\,,\\
\mathcal{L}_{eff} = &\frac{\mathrm{e}^{\chi/2}}{z^{d-3}}{A_{t}'}^2-\frac{\mathrm{e}^{-\chi/2}}{z^{d+1}}(2 \Lambda+d(d+1)f)\\
&+\frac{\mathrm{e}^{-\chi/2}}{2z^{d-1}}(-2f''+3f'\chi'+f({\chi'}^2+2\chi''))-(d-1)(d-2)\frac{\mathrm{e}^{-\chi/2}}{z^{d-1}}f{\zeta'}^2\\&+\frac{\mathrm{e}^{\chi/2}q^2 A_{t}^2 \rho_{x}^2}{\mathrm{e}^{2(d-2)\zeta}f z^{d-3}}+\frac{\mathrm{e}^{-\chi/2} f{\rho_{x}'}^2}{\mathrm{e}^{2(d-2)\zeta}z^{d-3}}+ \frac{d\mathrm{e}^{-\chi/2}}{z^d}(2f'-f\chi')\,,
\end{aligned}
\end{equation}
with the ``volume" $\mathcal{V}_d=\int dtdxd^{d-2}y$. The effective action depends on the derivatives of ($f$, $\chi$) up to second order and the derivatives of ($\zeta$, $\rho_{x}$, $A_{t}$) up to first order.\,\footnote{The effective action~\eqref{EffectAction} does not include the Gibbons-Hawking boundary term which has nothing to do with the bulk equations of motion for constructing the radially conserved quantity $\mathcal{Q}(z)$ below.} There is a scaling symmetry associated with $\mathcal{S}_{eff}$.
\begin{equation}\label{ScalingSymmetry}
\begin{aligned}
z\to \lambda z\,,\quad h\to h\,,\quad \chi\to \chi-2d\ln{\lambda}\,,\\
\zeta\to \zeta+\frac{\ln{\lambda}}{d-2}\,,\quad A_{t}\to \lambda^{d-1}A_{t}\,,\quad \rho_{x}\to \rho_{x}\,,
\end{aligned}
\end{equation}
where $\lambda$ is a positive constant. Considering an infinitesimal transformation of~\eqref{ScalingSymmetry}, with Einstein equation, one can obtain the Noether charge~\cite{Cai:2021obq}
\begin{equation}\label{ConserveCharge}
\mathcal{Q}(z)=\frac{\mathrm{e}^{\chi/2}}{z^{d-1}}((f\mathrm{e}^{-\chi})'-z^2A_{t}'A_t+2f\zeta'\mathrm{e}^{-\chi})\,,
\end{equation}
for which $\mathcal{Q}'(z)=0$. Let us assume that there exists an inner Cauchy horizon located at $z_I$ behind the event horizon $z=z_H$. Then we have
\begin{equation}
f(z_H)=f(z_I)=0, 
\quad f'(z_H)<0, \quad f'(z_I)\ge0,\quad z_H < z_I\,,
\end{equation} 
in our coordinate system.\,\footnote{We limit ourselves to the hairy black hole at finite temperature for which $f(z_H)<0$.}
Besides, the smoothness of both 
metric and matter fields near the horizon implies
\begin{equation}
A_{t}(z_H)=A_{t}(z_I)=0\,.
\end{equation} 
Computing $\mathcal{Q}$ at both event and inner horizons, we obtain
\begin{equation}\label{HorizonConserveCharge}
\frac{\mathrm{e}^{-\chi(z_H)/2}}{{z_H}^{d-1}}f'(z_H)=\frac{\mathrm{e}^{-\chi(z_I)/2}}{{z_I}^{d-1}}f'(z_I)\,.
\end{equation}
One now finds that the left hand side of~\eqref{HorizonConserveCharge} is negative, while the right hand side is non-negative. Therefore, the equation~\eqref{HorizonConserveCharge} cannot be satisfied and a smooth inner horizon is never able to form. More recently, the no inner horizon results were shown for the model with three real massive vector fields~\cite{DeClerck:2023fax}. It can only work for the case where the mass squared of each vector field is negative. Our method applies to such massive vector case independent of the sign of the mass squared.

In the absence of an inner horizon, there will be a cosmological interior evolution towards a spacelike singularity at $z\rightarrow\infty$. Before reaching the singularity, several intermediate regimes develop. Our interest here is the subsequent evolution in the far interior, \emph{i.e.} the alternation of Kasner epochs.

\section{Kasner Structure and Transformation Law}\label{SubSec:KasStructure}
Since the equations of motion~\eqref{EoM:zeta}-\eqref{EoM:h} are highly nonlinear, we are not able to obtain the solutions analytically. Motivated by the method developed in our previous work~\cite{Cai:2023igv}, one is able to obtain self-consistent asymptotic solutions under certain approximations. We take a Kasner epoch as starting point and consider possible deviation that leads to the alternation to another Kasner epoch.
Meanwhile, the results will be further established by checking the full numerical solutions.

\subsection{Kasner Solution}
Comparing to the equations of motion for scalar case~\cite{Cai:2023igv}, one finds that the metric function $\zeta$ plays a similar role of the scalar field. The Kasner epoch can be obtained by dropping terms that are shown to be negligible from our numerical solutions. More precisely, for $d\geq3$, the approximate equations of~$\eqref{EoM:zeta}$-$\eqref{EoM:h}$ in the far interior can be obtained as:
\begin{equation}\label{ApproximateEoMs}
\begin{aligned}
\zeta''=-\frac{1}{z}\zeta'&,\quad \chi'=2(d-2)z\zeta'^2,\quad h'=\frac{1}{2(d-1)}\left(\frac{\mathrm{e}^{\chi/2}A_{t}'}{z^{d-3}}\right)^2z^{d-3}\mathrm{e}^{-\chi/2}\,,\\
\quad &(z\rho_{x}')'=2\left((d-2)\zeta'-\frac{1}{z}\right)z\rho_{x}'\,,\quad \left(\frac{\mathrm{e}^{\chi/2}A_{t}'}{z^{d-3}}\right)'=0\,.
\end{aligned}
\end{equation}
Here we have assumed that $h'$ is integrable and the other terms are negligible. Solving the differential equations~\eqref{ApproximateEoMs}, one can obtain analytically that
\begin{equation}\label{ApproximateSol}
\begin{aligned}
\zeta=\beta\ln z&+C_{\zeta}, \quad z\rho_{x}'=C_{z\rho_{x}'}z^{2(d-2)\beta-2},\quad A_{t}'=C_{A_{t}'}z^{d-3}\mathrm{e}^{-\chi/2}\,,\\
&\chi=2(d-2)\beta^2\ln z+C_{\chi},\quad h'=C_{h'} z^{d-3}\mathrm{e}^{-\chi/2}\,,
\end{aligned}
\end{equation}
where $C_{\zeta}$, $C_{z\rho_{x}'}$, $C_{A_{t}'}$, $C_{\chi}$, $C_{h'}$ and $\beta$ are all constants. In particular, one has $C_{h'}>0$ from the third equation of~\eqref{ApproximateEoMs}. Then the metric reads
\begin{equation}\label{ApproximateMetric}
\mathrm{d}s^2 = C_t z^{(d-2)(1-\beta^2)}\mathrm{d}t^2 - \frac{ \mathrm{d}z^2 }{C_z z^{(d+2)+(d-2)\beta^2}}+ C_x z^{2(d-2)\beta-2}\mathrm{d}x^2 + \frac{\mathrm{d}\Sigma^2_{d-2}}{C_{\Sigma}z^{2\beta+2}}\,.
\end{equation}
One then finds that all metric components are power laws and $\zeta$ is logarithmic. According to the ``No inner horizon theorem'' proved in Section~\ref{SubSec:Noether Charge}, one knows that $C_{t}$, $C_{x}$, $C_{z}$ and $C_{\Sigma}$ are positive constants. Converting to the proper time $\tau\sim z^{-\frac{d}{2}-\frac{(d-2)\beta^2}{2}}$, we have
\begin{equation}\label{eqKas}
\begin{split}
\mathrm{d}s^2=-\mathrm{d}\tau^2+c_t \tau^{2p_t}\mathrm{d}t^2+c_x\tau^{2p_x}\mathrm{d}x^2+c_{\sigma}\tau^{2p_{\sigma}}\mathrm{d}\Sigma^2_{d-2}\,,
\end{split}
\end{equation}
with
\begin{equation}\label{Kbeta}
\begin{aligned}
p_t=\frac{(d-2)(\beta^2-1)}{(d-2)\beta^2+d},\quad p_x=\frac{2-2(d-2)\beta}{(d-2)\beta^2+d},\quad p_{\sigma}=\frac{2(\beta+1)}{(d-2)\beta^2+d}\,,
\end{aligned}
\end{equation}
and ($c_{t}$, $c_{x}$, $c_{\sigma}$) constants. Note that the proper time $\tau$ locates the singularity at $\tau=0$. This is a Kasner universe with the Kasner exponents $p_{t}$, $p_{x}$ and $p_{\sigma}$ satisfying
\begin{equation}
  p_t+p_x+(d-2)p_{\sigma}=1,\quad p_t^2+p_{x}^2+(d-2)p_{\sigma}^2=1\,.
\end{equation}
It is easy to show that in the Kasner universe~\eqref{eqKas} one metric component grows while others shrink for $d=3$ case, while at least one metric component grows for higher dimensions. We will see this feature more clearly in Sec.~\ref{SubSec:KasStructure}. Nevertheless, the overall spatial volume density collapses as approaching the singularity.

We emphasize that to obtain the Kasner universe, we have two requirements: $h'$ is integrable and many terms in the full equations of motion are negligible. Once they are invalid, the solution~\eqref{ApproximateSol} will become unstable towards the deep interior.
In fact, these two conditions define two important boundaries of dynamical process triggering the alternation of Kasner epochs. 

\subsection{Kasner Inversion}\label{subsec_inv}
A particularly simple case is caused by the non-integrability of $h'$. From~\eqref{ApproximateSol}, one has
\begin{equation}\label{hpp}
   h'(z)\sim z^{(d-3)-(d-2)\beta^2}\,.
\end{equation}
The no inner horizon theorem of Section~\ref{SubSec:Noether Charge} requires $h<0$ when $z>z_H$. If the integral of $h'$ diverges, $h$ will become positive at a particular point towards $z\rightarrow\infty$. Therefore, new dynamics should come into play, triggering the transformation to a different epoch.

From~\eqref{hpp}, the non-integrability of $h'$ yields
\begin{equation}\label{hcondition}
    (d-3)-(d-2)\beta^2>-1 \Rightarrow |\beta|<1\,,
\end{equation}
for which there is no stable Kasner epoch all the way down to the singularity. Actually, the term $h'/h$ under the condition~\eqref{hcondition} cannot be dropped and the approximate differential equations become
\begin{equation}\label{ApproximateInvDiff}
\zeta''=-\left(\frac{1}{z}+\frac{h'}{h}\right)\zeta',\quad \chi'= 2z(d-2)\zeta'^2,\quad h'=C_{h'} z^{d-3}\mathrm{e}^{-\chi/2}\,.
\end{equation}
The above equations can be solved using the
constant variant method, for which we take 
\begin{equation}\label{logbeta}
\zeta=\int^z \frac{\widetilde\beta(s)}{s}\mathrm{d}s\,.
\end{equation}
Substituting~\eqref{logbeta} into~\eqref{ApproximateInvDiff}, one obtains
\begin{equation}\label{InvDiffbeta}
z\widetilde\beta\widetilde\beta''-2z\widetilde\beta'^2+[(d-2)\widetilde\beta^2-(d-3)]\widetilde\beta'\widetilde\beta=0\,,
\end{equation}
which can be solved analytically via the following implicit function.
\begin{equation}\label{InvAlphaSol}
2(d-2)\ln\left[\frac{z}{z_{I}}\right]+\frac{2c_1}{\sqrt{c_1^2-1}}\mathrm{arctanh}\left[\frac{c_1-\widetilde\beta}{\sqrt{c_1^2-1}}\right]+\ln\left|\frac{(c_1^2-1)\widetilde\beta^2}{c_1^2(\widetilde\beta^2-2c_1\widetilde\beta+1)}\right|=0\,,
\end{equation}
where $c_1$ and $z_I$ are two constants. To better understand this solution, we introduce $z_I$ that satisfies $\widetilde{\beta}[z_I] = c_1$. It is easy to show that $z_I$ locates at the region where $\widetilde\beta$ changes abruptly. Moreover, far away from $z_I$, $\widetilde\beta$ becomes almost a constant, corresponding to a Kasner epoch described by~\eqref{ApproximateSol}. The value of the Kasner exponent $\beta$ before or after the transformation is obtained by taking the limit $z/z_I \ll 1$ or $z/z_I \gg 1$. We call this kind of Kasner alternation as Kasner inversion as it is reminiscent of the one in the scalar case~\cite{Cai:2023igv}.

The relation between the two Kasner exponents can be obtained as follows. Note that both the arctanh term and the last term of~\eqref{InvAlphaSol} go to infinity at the same time, because
\begin{equation}
   \left|\frac{c_1-\widetilde\beta}{\sqrt{c_1^2-1}}\right|\rightarrow 1\quad \Leftrightarrow \quad \widetilde\beta^2-2c_1\widetilde\beta+1\rightarrow 0\,.
\end{equation}
Therefore, when the first term of~\eqref{InvAlphaSol} tends to infinity (\emph{i.e.} far away from $z_I$), to maintain the validity of~\eqref{InvAlphaSol}, the last two term of~\eqref{InvAlphaSol} should tend to infinity as offset. When $z$ goes from $z/z_{I}\ll 1$ to $z/z_{I}\gg 1$, $\ln[z/z_{I}]$ changes from $-\infty$ to $+\infty$, suggesting that the last two terms should go from $+\infty$ to $-\infty$. Consequently, the two exponents $\beta$ for the Kasner epochs before and after the Kasner inversion are the \textbf{roots} of the following quadratic equation for $\widetilde\beta$:
\begin{equation}
    \widetilde\beta^2-2 c_1\widetilde\beta+1=0\,.
\end{equation}
According to Vieta's formula for quadratic equation, we can obtain the transformation law between two adjacent Kasner epochs.
\begin{equation}\label{KasInvLaw}
  \beta\beta_{I}=1\,,
\end{equation}
where $\beta$ is the Kasner exponent before the Kasner inversion, and $\beta_{I}$ is the one after the inversion. Interestingly, unlike the case with scalar hair~\cite{Cai:2023igv}, the transformation law of Kasner inversion is independent of the spacetime dimension.

\subsection{Kasner Transition and Reflection}\label{sec:KR}

We return to the case where $h'$ is integrable and therefore $h'/h$ can be dropped from~\eqref{EoM:zeta}. Then the approximate equation for $\zeta$ becomes
\begin{equation}\label{ApproximateZeta}
\zeta''=-\frac{1}{z}\zeta'-\frac{1}{d-1}(z\rho_{x}')^2\mathrm{e}^{-2(d-2)\zeta}+\frac{q^2A_{t}^2\rho_{x}^2}{(d-1)z^{2d-2}h^2}\mathrm{e}^{-2(d-2)\zeta}\,.
\end{equation}
Depending on whether $\rho_{x}'$ is integrable or not, we have the following two cases.

\paragraph{For non-integrable $\rho_{x}'$} When $\rho_{x}'$ is non-integrable, it behaves approximately as a polynomial and then one can take $\rho_{x}\sim z\rho_{x}'$ as a good approximation. Meanwhile, both $h$ and $A_t$ are finite. Therefore, we can drop the last term of~\eqref{EoM:rho}. The resulting the approximate equation for $\rho_{x}$ is
\begin{equation}\label{ApproximateRho}
(z\rho_{x}')' = 2\left((d-2)\zeta'-\frac{1}{z}\right)z\rho_{x}'\,.
\end{equation}
Integrating the above equation yields
\begin{equation}\label{SolutionRho}
z\rho_{x}'=\rho_0\frac{\mathrm{e}^{2(d-2)\zeta}}{z^2}\,,
\end{equation}
with $\rho_0$ a constant. From~\eqref{SolutionRho}, we can derive that the the non-integrable condition of $\rho_{x}'$ is
\begin{equation}
\beta>\frac{1}{d-2}\,,
\end{equation}
where we have assumed that $\zeta\sim\beta \ln z$ is a good approximation.

We now check the order of the last term we have dropped from~\eqref{EoM:rho}. Considering that both $A_t$ and $h$ are finite, we have
\begin{equation}\label{ignore}
    \mathcal{O}\left(\frac{q^2A_t^2\rho_{x}}{z^{2d-1}h^2}\right)=\mathcal{O}\left(\frac{\rho_{x}}{z^{2d-1}}\right)<\mathcal{O}\left(\frac{\rho_{x}}{z}\right)=\mathcal{O}\left(\frac{z\rho_{x'}}{z}\right)\,.
\end{equation}
In the last equality we take $\rho_{x}\sim z\rho_{x}'$ as a good approximation since $\rho_{x}'$ is non-integrable and polynomial. Therefore, we can safely drop the last from~\eqref{EoM:rho} in the present case.

\paragraph{For integrable $\rho_{x}'$} When $\rho_{x}'$ is integrable, $\rho_{x}$ will be upper bounded. Then we have from~\eqref{EoM:rho} that
\begin{equation}
\mathcal{O}\left((z\rho_{x}')'\right)\sim \mathcal{O}\left(\left((d-2)\zeta'-\frac{1}{z}\right)z\rho_{x}'\right)\sim\mathcal{O}\left(\rho_{x}'\right).
\end{equation}
If the last term in \eqref{EoM:rho} remains negligible, the solution~\eqref{SolutionRho} of $z\rho_{x}'$ still holds. However, when it becomes comparable to other terms, from~\eqref{EoM:rho}, we can derive that
\begin{equation}
\mathcal{O}\left(\rho_{x}'\right)\sim \mathcal{O}\left(\frac{q^2A_t^2\rho_{x}}{z^{2d-1}h^2}\right)=\mathcal{O}\left(\frac{1}{z^{2d-1}}\right)\,,
\end{equation}
since in this case $A_t$, $h$ and $\rho_x$ are finite. Therefore, when $\beta<1/(d-2)$, we are able to estimate the order of \(\rho_{x}'\):
\begin{equation}
\mathcal{O}\left(\rho_{x}'\right)=\mathrm{max}\left\{\mathcal{O}\left(\frac{\mathrm{e}^{2(d-2)\zeta}}{z^3}\right), \;\;\mathcal{O}\left(\frac{q^2A_t^2\rho_{x}}{z^{2d-1}h^2}\right)\right\}.
\end{equation}
When $\beta<-1$, it is easy to verify that 
\begin{equation}
\mathcal{O}\left(\frac{\mathrm{e}^{2(d-2)\zeta}}{z^3}\right)<\mathcal{O}\left(\frac{q^2A_t^2\rho_{x}}{z^{2d-1}h^2}\right).
\end{equation}
To summarize, we can obtain the following properties about $\rho_{x}'$:
\begin{equation}\label{RhoAnalysis}
\begin{cases}
\mathrm{Non\ Integrable}: \mathcal{O}\left(\rho_{x}'\right)=\mathcal{O}\left(\frac{\mathrm{e}^{2(d-2)\zeta}}{z^3}\right),\quad \beta>\frac{1}{d-2}\,, \\
\,\,\mathrm{Integrable}: 
\begin{cases}
\quad \mathcal{O}\left(\rho_{x}'\right)=\mathcal{O}\left(\frac{\mathrm{e}^{2(d-2)\zeta}}{z^3}\right),\quad -1<\beta<\frac{1}{d-2}\,,\\
\quad \mathcal{O}\left(\rho_{x}'\right)=\mathcal{O}\left(\frac{1}{z^{2d-1}}\right),\qquad \beta<-1\,.\\
\end{cases}
\end{cases}
\end{equation}
We now turn our attention to~\eqref{ApproximateZeta}.  After substituting~\eqref{RhoAnalysis} into~\eqref{ApproximateZeta}, we find three different cases when solving the resulted equation.

\paragraph{Case I: $\beta>1/(d-2)$} For this case, $\rho_{x}'$ is not integrable. With the property in the first line of ~\eqref{RhoAnalysis}, the order of last two term in the right hand of~\eqref{ApproximateZeta} can be estimated as
\begin{equation}\label{TransitionOrderAnalysis1}
\begin{aligned}
\mathcal{O}\left((z\rho_{x}')^2\mathrm{e}^{-2(d-2)\zeta}\right)&=\mathcal{O}\left(\frac{\mathrm{e}^{2(d-2)\zeta}}{z^4}\right)>\mathcal{O}\left(\frac{1}{z^2}\right),\\
\quad\mathcal{O}\left(\frac{q^2\rho_{x}^2}{z^{2d-2}}\mathrm{e}^{-2(d-2)\zeta}\right)&=\mathcal{O}\left(\frac{\mathrm{e}^{2(d-2)\zeta}}{z^{2d+2}}\right)<\mathcal{O}\left(\frac{\mathrm{e}^{2(d-2)\zeta}}{z^4}\right).
\end{aligned}
\end{equation}
It is obviously that the first term of~\eqref{TransitionOrderAnalysis1} is dominant as $\beta>1/(d-2)$. 
Thus, substituting~\eqref{SolutionRho} into~\eqref{ApproximateZeta}, we obtain
\begin{equation}\label{ApproximateTranDiff1}
\frac{1}{z}(z\zeta')'\simeq \frac{\mathrm{e}^{2(d-2)\zeta}}{z^4}\,.
\end{equation}
Using the constant variant method as~\eqref{logbeta}, we have the following  differential equation for $\widetilde\beta$:
\begin{equation}\label{TranDiffbeta1}
z \widetilde{\beta}''+3\widetilde\beta'-2(d-2)\widetilde\beta'\widetilde\beta=0\,.
\end{equation}
Integrating the above equation yields
\begin{equation}
\widetilde{\beta}(z)=\frac{1}{d-2}-c_1 \tanh\left[c_1\ln(z/z_T)\right]\,,
\end{equation}
where $c_1$ and $z_T$ are integration constants.  Far away from $z_T$, $\widetilde\beta$ becomes a constant, corresponding to a Kasner epoch described by~\eqref{ApproximateSol}. The alternation of the two Kasner epochs happens near $z_T$. 

Denoting $\beta$ and $\beta_T$ as the Kasner exponents before and after the alternation, respectively, it is easy to obtain that 
\begin{equation}\label{KT}
\beta+\beta_{T}=\frac{2}{d-2}\,.
\end{equation}
The above law is reminiscent of the Kasner transition in the scalar case~\cite{Cai:2023igv}. Therefore, we also call it Kasner transition and it depends on the spacetime dimension.

One should pay more attention when $1/(d-2)<\beta<1$ for which there develops an overlapping region between the Kasner inversion and transition, and there may be competing
behaviors between the Kasner inversion and transition in the overlapping region. To solve this issue, we come back to~\eqref{EoM:zeta} and estimate the contribution of each term. In particular, we need to compare the following two terms: $(z\rho_{x}')^2\mathrm{e}^{-2(d-2)\zeta}$ and $h'\zeta'/h$. We then find that there is a critical Kasner exponent 
\begin{equation}
	\beta_c=-1+\sqrt{\frac{2(d-1)}{d-2}}\,,
\end{equation}
for which
\begin{equation}
	\mathcal{O}\left((z\rho_{x}')^2\mathrm{e}^{-2(d-2)\zeta}\right)=\mathcal{O}\left(\frac{\mathrm{e}^{2(d-2)\zeta}}{z^4}\right)=\mathcal{O}\left(\frac{h'\zeta'}{h}\right)\,.
\end{equation}
Therefore, when $\beta>\beta_c$, the term $(z\rho_{x}')^2\mathrm{e}^{-2(d-2)\zeta}$ regarding the Kasner transition dominates the dynamics. In contrast, when $\beta<\beta_c$ the Kasner alternation is governed by the Kasner inversion.

\paragraph{Case II: $-1<\beta<1/(d-2)$} Considering the second line of ~\eqref{RhoAnalysis}, we can estimate the order of last two terms in the right hand of~\eqref{ApproximateZeta}.
\begin{equation}\label{TransitionOrderAnalysis2}
\mathcal{O}\left((z\rho_{x}')^2\mathrm{e}^{-2(d-2)\zeta}\right)<\mathcal{O}\left(\frac{1}{z^2}\right),\quad\mathcal{O}\left(\frac{\rho_{x}^2}{z^{2d-2}}\mathrm{e}^{-2(d-2)\zeta}\right)<\mathcal{O}\left(\frac{1}{z^{2}}\right)\,.
\end{equation}
Therefore, both terms are negligible. Hence, combined with the above analysis in subsection~\ref{subsec_inv}, Kasner inverison will dominate the transformation process in this $\beta$ region.

\paragraph{Case III: $\beta<-1$} For this case, $\rho_{x}'$ is integrable and $\rho_{x}$ has an upper bound. The order of last two terms in the right hand of~\eqref{ApproximateZeta} can be estimated as
\begin{equation}\label{TransitionOrderAnalysis3}
\begin{aligned}
\mathcal{O}\left((z\rho_{x}')^2\mathrm{e}^{-2(d-2)\zeta}\right)&=\mathcal{O}\left(\frac{\mathrm{e}^{-2(d-2)\zeta}}{z^{2(2d-2)}}\right)<\mathcal{O}\left(\frac{\rho_{x}^2}{z^{2d-2}}\mathrm{e}^{-2(d-2)\zeta}\right),\\
\quad\mathcal{O}\left(\frac{q^2\rho_{x}^2}{z^{2d-2}}\mathrm{e}^{-2(d-2)\zeta}\right)&=\mathcal{O}\left(\frac{\mathrm{e}^{-2(d-2)\zeta}}{z^{2d-2}}\right)>\mathcal{O}\left(\frac{1}{z^2}\right).
\end{aligned}
\end{equation}
Clearly, the last one of~\eqref{TransitionOrderAnalysis3} is dominant when $\beta<-1$. Therefore, we can obtain the approximate equation about $\zeta$:
\begin{equation}\label{ApproximateTranDiff2}
\frac{1}{z}(z\zeta')'\simeq \frac{\mathrm{e}^{-2(d-2)\zeta}}{z^{2d-2}}\,.
\end{equation}
Substituting~\eqref{logbeta} into the above equation, we obtain
\begin{equation}\label{TranDiffbeta2}
z \widetilde{\beta}''+(2d-3)\widetilde\beta'+2(d-2)\widetilde\beta'\widetilde\beta=0\,,
\end{equation}
which can be solved analytically.
\begin{equation}\label{RefAlphaSol}
\widetilde{\beta}(z)=-1-c_2 \tanh\left[c_2\ln(z/z_R)\right]\,,
\end{equation}
with $c_2$ and $z_R$ integration constants.
Similar to the Kasner transition, $\widetilde\beta$ becomes a constant far away from $z_R$, corresponding to a Kasner epoch described by~\eqref{ApproximateSol}. The alternation of the two Kasner epochs is triggered near $z_R$. It is easy to obtain the relation between the two adjacent Kasner epochs:
\begin{equation}~\label{KR}
\beta+\beta_{R}=-2\,,
\end{equation}
where $\beta$ is the Kasner exponent before the alternation, and $\beta_{R}$ is the one after the alternation. In contrast to the Kasner transition of~\eqref{KT}, the law of~\eqref{KR} is independent of the dimension of the spacetime. We shall call it Kasner reflection whose meaning will be clear in the next section.

\subsection{Laws of Kasner Transformation}
In summary, we find that the following laws describing the alternation of adjacent Kasner  epochs inside anisotropic black hole with vector hair.
\begin{equation}\label{KasLawAll}
\begin{cases}
\mathrm{Kasner\ transition}: \beta+\beta_T=\frac{2}{d-2},\quad \beta>\beta_c\,, \\
\mathrm{Kasner\ inversion}: \quad \beta\,\beta_{I}=1,\quad -1<\beta<\beta_c\,,\\
\mathrm{Kasner\ reflection}: \beta+\beta_R=-2,\quad \beta<-1\,,\\
\end{cases}
\end{equation}
where $\beta_{c}=-1+\sqrt{2(d-1)/(d-2)}$. As clearly from~\eqref{Kbeta}, all Kasner exponents of a Kasner universe are fully characterized by the parameter $\beta$. One can see from~\eqref{KasLawAll} that our laws cover all parameter space of $\beta$, suggesting that we have clarified all possible kinds of Kasner transformation for the anisotropic black holes with vector hair~\eqref{GenAnsatz}.

Interestingly, we find that while the law of Kasner transition depends on the dimension of spacetime, the laws of Kasner inversion and Kasner reflection are dimension-independent. In particular, when $d=3$, there is a symmetric region between the Kasner reflection and transition, for which the transformation laws~\eqref{KasLawAll} are completely same with the situation of pure gravity~\cite{Lifshitz:1963ps,Belinsky:1970ew} and scalarized case~\cite{An:2022lvo}. The transformation laws~\eqref{KasLawAll} suggest the chaotic evolution characterized by the random change of the amplitude of the Kasner exponent at late interior times. In particular, independent of the spacetime dimension, there exist always the fixed point $\beta=-1$ approaching which numerous Kasner epochs will be triggered with the Kasner exponent $\beta$ been amplified continuously towards the singularity (see Section~\ref{sec:koscillation}).

Besides, we find that the Kasner transformation and the minimum negative value of the Kasner exponents is closely correlated. Considering the expressions of Kasner exponents~\eqref{Kbeta}, we can obtain 
\begin{equation}\label{betaDistribution}
\mathrm{Min}\{p_t,p_x,p_{\sigma}\}=\begin{cases}
p_x<0,\ \mathrm{Kasner\ transition}:\beta>\beta_c\,, \\
p_t<0,\ \mathrm{Kasner\ inversion}:  -1<\beta<\beta_c\,,\\
p_{\sigma}<0,\ \mathrm{Kasner\ reflection}: \beta<-1\,.\\
\end{cases}
\end{equation}
It suggests that, in different $\beta$ regions, the corresponding minimum and negative Kasner exponent leads to the instability of Kasner space-time, and then the Kasner transformation.

It is interesting to compare our approach to the the cosmological billiard method in terms of an elegant geometric description. The basic elements of the latter is to know the specific locations of the walls (\emph{i.e.} the boundaries of the billiard table), and the reflection rules about the walls. The corresponding transformation rules for the Kasner exponent have been worked out in general (without symmetry assumptions) in~\cite{Demaret:1986ys,DEMARET198527} for gravitational walls and in~\cite{Damour:2000wm,Damour:2002fz} for p-form walls. In $d=3$, the gravitational and the electromagnetic walls coincide and there is no symmetry wall when the metric is diagonal. Therefore, the billiard is the ideal triangle delimited by the gravitational walls. The collision against a gravitational wall yields as the following Kasner exponents~\cite{Damour:2002et}
\begin{equation}\label{kastra3d}
    p_1^{\prime}=-\frac{p_1}{1+2p_1},\quad p_2^{\prime}=\frac{p_2+2p_1}{1+2p_1},\quad p_3^{\prime}=\frac{p_3+2p_1}{1+2p_1}\,,
\end{equation}
where $\{p_1,p_2,p_3\}$ denote the Kasner exponents with different spatial directions and $p_1$ is the smallest Kasner exponent. Note that $\beta_c=1$ for $d=3$. From~\eqref{betaDistribution}, one has $p_1=p_x$ when $\beta>1$, for which the formula~\eqref{kastra3d} gives our Kasner transition formula. When $|\beta|<1$, it is $p_t$ that is the smallest exponent (\emph{i.e.} $p_1=p_t$) and the formula~\eqref{kastra3d} gives then the Kasner inversion formula. Finally, when $\beta<-1$, it is $p_\sigma$ that is the smallest exponent (\emph{i.e.} $p_1=p_\sigma$), for which the formula~\eqref{kastra3d} reduces to the Kasner reflection. This demonstrates the truth of our formula~\eqref{KasLawAll} and suggests that our theory could be described by cosmological billiard.

Regarding to our present theory, to the best of our knowledge, there has no explicit formula as~\eqref{kastra3d} via cosmological billiard when $d>3$ in the literature. Indeed, we find that, when $d>3$, our laws of Kasner transformation~\eqref{kastra3d} can be summarized as 
\begin{equation}\label{kastrad}
     p_1^{\prime}=-\frac{p_1}{1+\frac{n_1^2}{d-2}2p_1},\quad p_2^{\prime}=\frac{p_2+\frac{n_1^2}{d-2}2p_1}{1+\frac{n_1^2}{d-2}2p_1},\quad p_3^{\prime}=\frac{p_3+\frac{n_1^2}{d-2}2p_1}{1+\frac{n_1^2}{d-2}2p_1}\,,
\end{equation}
with $p_1< p_2< p_3$.  Here $n_{1}$ denotes the coefficient of $p_{1}$ in the Kasner relationship $p_{t}+p_{x}+(d-2)p_{\sigma}=1$. More precisely, $n_{1}=1$ when $p_{1}=p_{t}$ or $p_{1}=p_{x}$, and $n_1=d-2$ when $p_{1}=p_{\sigma}$. It is interesting to understand the interior dynamics using cosmological billiards and to check if the surviving various types of walls yield the formula we obtain here. Nevertheless, it is beyond the scope of our present study. While we believe that the cosmological billiards method can also handle the Kasner dynamics in the present model, our approach is easy to generalize to more general interactions.

Before ending this section, we should check the consistency of our analytic treatment, since we have dropped some terms from the full equations of motion~\eqref{EoM:zeta}-\eqref{EoM:h}. More precisely, we should check if these terms we discarded are small in a given Kasner universe and Kasner transformation. We begin with the cosmological constant $\Lambda$ which appears in~\eqref{EoM:h}. We have 
\begin{equation}
\mathcal{O}\left(\frac{2\Lambda}{z^{d+1}}\right)=\mathcal{O}\left(\frac{1}{z^{d+1}}\right),
\end{equation}
which is integrable for $d \geq 3$, thus it can be consistently neglected. In other words, the asymptotic behavior of spacetime has no robust effect on the internal structure of black holes. 
Let's then consider another term regarding $\rho_{x}$. In the above analysis, we have assumed that the right side of~\eqref{EoM:At} is integrable. However, this assumption is only necessary during the Kasner inversion. In the Kasner transition and Kasner reflection cases, even if this assumption is violated, the assumption that $h'$ is integrable will remain valid.

\textbf{When $|\beta|<1$}, with~\eqref{RhoAnalysis}, it is easy to verify that
\begin{equation}
    \mathcal{O}\left(\frac{2q^2A_{t}\rho_{x}^2}{z^{2d-3}h}\mathrm{e}^{-2(d-2)\zeta}\right)=\mathcal{O}\left(\frac{\rho_{x}^2}{z^{2d-3}}\mathrm{e}^{-2(d-2)\zeta}\right)<\mathcal{O}\left(\frac{1}{z}\right),
\end{equation}
with $\zeta\sim\beta \ln z$ a good approximation. Therefore, assuming the right side of~\eqref{EoM:At} to be integrable is self-consistent in the Kasner inversion process. 

\textbf{When $|\beta|>1$}, if the term in the right side of~\eqref{EoM:At} is not integrable, we could assume 
\begin{equation}
   \frac{2q^2A_{t}\rho_{x}^2}{z^{2d-3}h}\mathrm{e}^{-2(d-2)\zeta}\sim z^{n-1}, \quad n\geq 0\,.
\end{equation}
At this point, the evolution of $\chi$ is dominated by the last term of~\eqref{EoM:chi}, from which we can obtain that
\begin{equation}    \mathcal{O}\left(\chi'\right)=\mathcal{O}\left(z^{n-1}\right)\,.
\end{equation}
Therefore, we can estimate the order of the last term in~\eqref{EoM:h} and find that
\begin{equation}\label{estimateh}    \mathcal{O}\left(\left(\frac{\mathrm{e}^{\chi/2}A_{t}'}{z^{d-3}}\right)^2z^{d-3}\mathrm{e}^{-\chi/2}\right)=\mathcal{O}\left(z^{2n+d}\mathrm{e}^{-\lambda z^n}\right),
\end{equation}
where $\lambda$ is a positive constant. For any finite value of $d$, the left side of~\eqref{estimateh} is obviously integrable. Thus, it is self-consistent to assume that $h'$ is integrable when considering the Kasner transition and reflection analysis. It should be noted that, the approximate form of equations~\eqref{ApproximateEoMs} regarding $\rho_{x}$ will change to
\begin{equation}\label{ApproximateEoMs2}
\rho_{x}'\sim \mathrm{max}\left\{\mathcal{O}\left(\frac{\mathrm{e}^{2(d-2)\zeta}}{z^3}\right), \mathcal{O}\left(\frac{1}{z^{2d-1}}\right)\right\},
\end{equation}
which still gives a Kasner universe because $h'$ is integrable and the terms leading to Kasner inversion is negligible. 

Thus far, we have analytically checked the self-consistency of all the assumptions when obtaining the laws of Kasner transformation~\eqref{KasLawAll}. Under certain approximations, we can obtain self-consistent asymptotic solutions. In the next part, our analytically approach will be further established by comparing to the full numerical solutions.

\subsection{Numerical Check}\label{sec:Ncheck}

To numerically solve the coupled equations of motion~\eqref{EoM:zeta}-\eqref{EoM:h}, we should consider appropriate boundary conditions. We impose the regularity conditions at the event horizon $z=z_H$, which mean that all our functions
have finite values and admit a series expansion in terms of $(z-z_H)$. In particular, we have $f(z_H)=A_t(z_H)=0$, and the temperature of the black hole reads
\begin{equation}
T=-\frac{e^{-\chi(z_H)}f'(z_H)}{4\pi}\,. 
\end{equation}
On the other hand, we have the general falloff near the AdS boundary at $z=0$.
\begin{equation}
\begin{split}
\rho_x=\rho_s+\cdots+\rho_v z^{d-2}+\cdots\,, \quad A_t=\mu+\cdots\,,\\
f=1+\cdots\,,\quad \chi=0+\cdots\,,\quad\zeta=0+\cdots\,,
\end{split}
\end{equation}
with the dots standing for the higher order terms in the expansion of $z$. For definiteness, we limit ourselves to the case for which the vector hair develops spontaneously, \emph{i.e.} $\rho_s=0$. It corresponds to the holographic p-wave superconductor~\cite{Cai:2013pda,Cai:2013aca} where $\rho_v$ is the condensate and $\mu$ is the chemical potential. Without loss of generality, we shall fix the chemical potential to be $\mu=1$, which means all physical quantities are measured in unit of the chemical potential. Finally, we are left with only one-dimensional parameter space of solutions to explore, which can be labelled by the temperature (or equivalently $T/\mu$). More precisely, below a critical temperature $T_c$, the vector condensate develops smoothly, breaking the U(1) symmetry as well as rotational symmetry spontaneously. For the massless vector of~\eqref{VectorModel}, the phase transition from the normal phase for which $\rho_x=0$ to the broken phase with non-trivial $\rho_x$ can be a second order transition or a first order one, depending on the the value of charge $q$. Please consult~\cite{Cai:2013aca,Cai:2014ija} for more details, in particulate the numerical strategy for constructing the hairy solutions.

\subsubsection{Case with $d=3$}
\begin{figure}[H]
\centering
\includegraphics[width=0.7\textwidth]{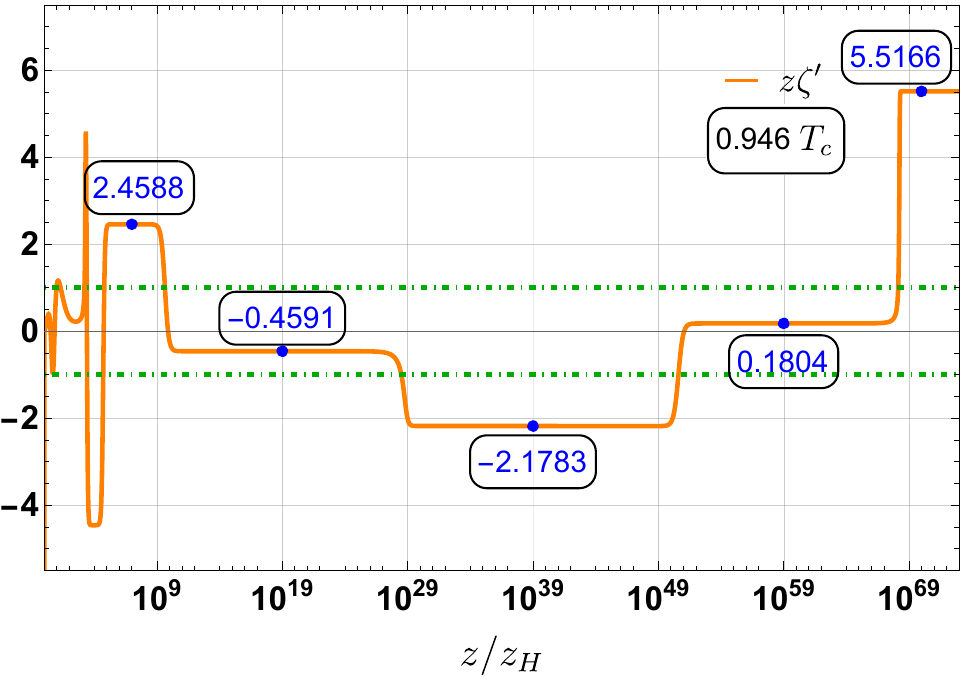}
\caption{An illustrative example of the black hole interior demonstrating function $z\zeta'$ for $d=3$, $q=\sqrt{2}$ and $L=1$. We numerically solve the complete equations of motion at $T\approx 0.946T_c$. Each plateau corresponds to a Kasner epoch with the number denoting the value of its Kasner exponent $\beta$. The two fixed points with $\beta=\{-1,1\}$ are denoted by the green-dashed lines. The validity of the transformation laws~\eqref{KasLawAll} for the alternation of Kasner epochs can be checked explicitly.}\label{4d_alternation}
\end{figure}

The interior dynamics with $d=3$ is presented in Figure~\ref{4d_alternation} for $z\zeta'$ from the horizon to a large radial $z$ value. One can see clearly that the far interior evolution shows a sequence of Kasner epochs (plateaus in Figure~\ref{4d_alternation}) connected by bounces. We present the value of the Kasner exponent $\beta$ in each Kasner epoch by fitting the numerical solutions, which in turn defines all the exponents of~\eqref{Kbeta}. As clearly from~\eqref{KasLawAll}, for $d=3$, the one-dimensional parameter space of $\beta$ for our transformation laws does not have any overlap. The two fixed points with $\beta=\{-1,1\}$ are represented by the green-dashed lines of Figure~\ref{4d_alternation}.
As shown in Figure~\ref{4d_alternation}, after the collapse and oscillation near the event horizon, around $z/z_H=10^{7}$ one has a clear Kanser epoch with $\beta_1=2.4588$ which is larger than the critical exponent $\beta_c=1$. Thus a Kasner transition is triggered and results in a Kasner epoch with $\beta_2=-0.4591$, well predicted by our analytical approach. Since $-1<\beta_2<1$, one should have the third Kasner epoch via the Kasner inversion. We numerically find that $\beta_3=-2.1783$ and thus $\beta_2 \beta_3=1.0000$ in good agreement with the Kasner inversion law~\eqref{KasLawAll}. The new Kasner epoch becomes unstable towards far interior and jumps to the fourth Kasner epoch via the Kasner reflection with the exponent $\beta_4=0.1804$. Then a second Kasner inversion triggers and the system arrivals at the fifth Kanser epoch with $\beta_5=5.5166$. Since $\beta_5>1$, the next Kasner alternation will be triggered via the Kasner transition. The process would repeat itself for ever as predicted by~\eqref{KasLawAll}.

\subsubsection{Case for $d>3$}
\begin{figure}[H]
\centering
\includegraphics[width=0.7\textwidth]{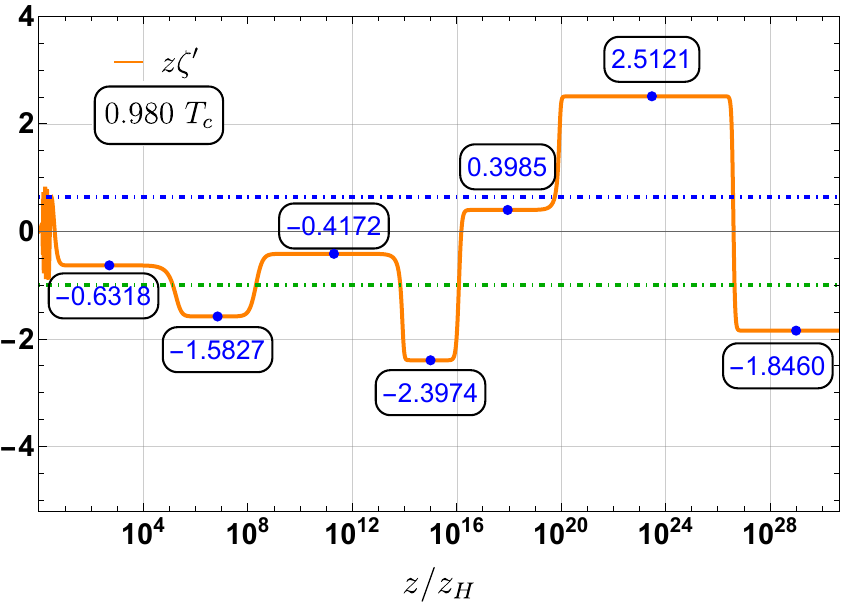}\;
\caption{The black hole interior demonstrating function $z\zeta'$ at $T\approx 0.982T_c$ (left) and $T\approx 0.975T_c$ (right). The Kasner exponent $\beta$ at each Kasner epoch is marked, from which one can verify the transformation laws~\eqref{KasLawAll} directly. The two fixed points with $\beta={-1,\beta_c(5)}$ are denoted by the green-dashed and blue-dashed lines. We have chosen $d=5$, $q=\sqrt{10}$ and $L=1$.}\label{6d_alternation}
\end{figure}

For the sake of specificity, we consider the case with $d=5$ for which the transformation law reads
\begin{equation}
\begin{cases}
\mathrm{Kasner\ transition}: \beta+\beta_T=\frac{2}{3},\quad \beta>\beta_c(5)\,, \\
\mathrm{Kasner\ inversion}: \quad \beta\,\beta_{I}=1,\quad -1<\beta<\beta_c(5)\,,\\
\mathrm{Kasner\ reflection}: \beta+\beta_R=-2,\quad \beta<-1\,,\\
\end{cases}
\end{equation}
where $\beta_{c}(5)\sim 0.63$. Our numerical computation shows the  develop of the next Kasner epoch with $\beta_4=-2.3974$. One can check that $\beta_4+\beta_5=-1.0011$, so the alternation is well described by the Kasner reflection. The seventh Kasner epoch with its component $\beta_5=0.3985$. The next Kasner epoch has the exponent $\beta_6=2.5121$. One then finds that $\beta_5\beta_6=1.0011$, so the alternation between the above two Kasner epochs is well captured by the Kasner inversion. The next Kasner epoch has the exponent $\beta_7=-1.8460$. One then finds that $\beta_6+\beta_7=0.6661\sim 2/3$, so the alternation between the above two Kasner epochs is well captured by the Kasner transition. As manifestly shown in Fig.~\ref{6d_alternation}, the numerical results are highly consistent with the transition patterns~\eqref{KasLawAll} we obtained. Our analysis suggests that there exhibits a never-ending oscillatory behavior towards the spacelike singularity.

\section{Explanation of Previous Numerical Observations}\label{Sec:old}
The interior of anisotropic black holes with vector hair was investigated in~\cite{Cai:2021obq} and~\cite{Sword:2022oyg}. While the collapse of the interior geometry and the rapid oscillations in vector condensate were clear understood in~\cite{Cai:2021obq}, the chaotic nature of the interior evolution characterized by endless alternating Kasner epochs was only reported numerically. In this section, we will provide a clear explanation of the alternation of Kasner epochs found numerically in~\cite{Cai:2021obq} and~\cite{Sword:2022oyg}.

\subsection{Alternation of Kasner Epochs}

In the earlier work presented in~\cite{Cai:2021obq}, it was discovered numerically that the interior of charged vector hairy black holes exhibits Kasner structure, and some rules of Kasner transformation were summarized. Here, we shall demonstrate that the empirical equations found  in~\cite{Cai:2021obq} exactly correspond to the Kasner inversion and Kasner transition laws~\eqref{KasLawAll} with $\beta>-1$.

The ansatz of~\cite{Cai:2021obq} reads
\begin{equation}\label{GenAnsatz0}
\mathrm{d}s^2 = \frac{1}{\hat{z}^2}\left(-\hat{f}(\hat{z}) e^{-\hat{\chi}(\hat{z})} \mathrm{d}t^2 + \frac{1}{\hat{f}(\hat{z})} \mathrm{d}\hat{z}^2 + u(\hat{z}) \mathrm{d}x^2 + \mathrm{d}\Sigma^2_{d-2} \right)\,,
\end{equation}
and seems to be different from the one we adopt in~\eqref{GenAnsatz}. Nevertheless, both are equivalent by taking the following coordinate transformation:
\begin{equation}\label{MetricTrans}
\begin{aligned}
\hat{z}&=z \mathrm{e}^{\zeta},\quad \hat{f}=f\left(1+2z\mathrm{e}^{\zeta'-\zeta}+z^2\mathrm{e}^{2(\zeta'-\zeta)}\right),\\
u&=\mathrm{e}^{2(d-1)\zeta},\quad \hat{\chi}=\chi-2\zeta+\ln\left(1+2z\mathrm{e}^{\zeta'-\zeta}+z^2\mathrm{e}^{2(\zeta'-\zeta)}\right).
\end{aligned}
\end{equation}
In the far interior, one has a simplified set of equations which allows analytic solutions with~\cite{Cai:2021obq}
\begin{equation}\label{PreSol}
    u\sim \hat{z}^{n_u},\quad \hat{A}_t~\sim \hat{z}^{n_A}\,,
\end{equation}
where $n_u$ and $n_A$ are two constants. Substituting~\eqref{MetricTrans} into~\eqref{PreSol}, one immediately obtains
\begin{equation}\label{NSol}
    n_u=\frac{4\beta}{1+\beta},\quad n_A=1-\beta\,,
\end{equation}
where we have used the asymptotic solution of $\zeta\sim\beta\ln z$. For the 4-dimensional case reported in~\cite{Cai:2021obq}, one can then easily verify that the two Kasner transformation laws they found are:
\begin{equation}\label{KasLawPre}
\begin{aligned}
\{n_u+\hat{n}_u=4,\, n_u<2\},\quad&\Leftrightarrow\quad  \{\beta\hat{\beta}=1, \, -1<\beta<1\}\,,\\
\{n_A+\hat{n}_A=0,\, 2<n_u<4\},\quad&\Leftrightarrow\quad \{\beta+\hat{\beta}=2,\, 1<\beta<\infty\}\,,
\end{aligned}
\end{equation}
where $\{n_u, n_A, \beta\}$ and $\{\hat{n}_u, \hat{n}_A, \hat{\beta}\}$ represent the exponents before and after a Kasner transformation. This is exactly the Kanser transformation laws described in~\eqref{KasLawAll} when $\beta>-1$. Moreover, as predicted by~\eqref{KasLawAll}, another 
transformation law 
\begin{equation}
\{n_A + \hat{n}_A = 4, \, n_u > 4\} \quad \Leftrightarrow \quad \{\beta + \hat{\beta} = -2\,, \, \beta < -1\},
\end{equation}
corresponding to the Kasner reflection should be found upon exploring the parameter space carefully. 
Indeed, the last case was later shown in some numerically examples in~\cite{Sword:2022oyg}.

\subsection{Kasner Oscillation}\label{sec:koscillation}
The interior of p-wave holographic superconductor was recently studied in~\cite{Sword:2022oyg} by considering the 4-dimensional Einstein-SU(2) Yang-Mills theory which is equivalent to the Einstein-Maxwell-vector theory in the present study, as shown explicitly in~\cite{Cai:2021obq}. The authors of~\cite{Sword:2022oyg} identified some temperatures near which numerical analysis demonstrated the appearance of a large number of alternating Kasner epochs. More precisely, the oscillations are centered around $z\chi'=2$ and the amplitude of oscillations grows towards the black hole singularity. Nevertheless, the mechanism of such oscillation-amplification behavior is unclear.

This interesting interior dynamics can be understood analytically by using our analytical laws~\eqref{KasLawAll}. It is straightforward to derive from the approximate equation \eqref{ApproximateEoMs} that
\begin{equation}
z\chi'=2{\tilde{\beta}}^2\,,
\label{TansLaw_chi}
\end{equation}
where we have used~\eqref{logbeta} and taken $d=3$.
Therefore, one can determine the value of $z\chi'$ through the value of the Kasner exponent $\beta$. We shall show that when the exponent $\beta$ approaches $-1$, an oscillation-amplification behavior around the fixed point $\beta=-1$ will occur. 

To simplify the analysis, let us introduce $\bar{\beta}=\beta+1$ for which the fixed point now is $\bar{\beta}=0$. The laws of Kasner inversion and Kasner reflection of~\eqref{KasLawAll} become
\begin{equation}\label{KasLawOsci}
\begin{aligned}
\mathrm{Kasner\ inversion}&: \quad \bar{\beta}_{I}=\frac{\bar{\beta}}{\bar{\beta}-1},\quad 0<\bar{\beta}<2\,,\\
\mathrm{Kasner\ reflection}&:\quad \bar{\beta}_R+\bar{\beta}=0,\quad \bar{\beta}<0\,,
\end{aligned}
\end{equation}
independent of the spacetime dimension. We begin with a Kasner epoch with its exponent $\bar{\beta}_{0}$ satisfying $0<\bar{\beta}_{0}<1$. According to~\eqref{KasLawOsci}, there is a second Kasner epoch after the Kasner inversion with the exponent $\bar{\beta}_{1}=\frac{\bar{\beta}_{0}}{\bar{\beta}_{0}-1}<0$. It is clear that $|\bar{\beta}_1|>\bar{\beta}_{0}$ because $|\bar{\beta}_1|-\bar{\beta}_0=\frac{\bar{\beta}_{0}^{2}}{1-\bar{\beta}_{0}}>0$. Then, the Kasner reflection process will be activated, leading to the Kasner epoch with $\bar{\beta}_2=-\bar{\beta}_1>0$. Therefore, we can establish the relationship of the exponent $\bar{\beta}$ in a series of Kasner transformations:
\begin{equation}\label{KasOsciAmp}
    |\bar{\beta}_0|<|\bar{\beta}_1|=|\bar{\beta}_2|<|\bar{\beta}_3|=|\bar{\beta}_4|<|\bar{\beta}_5|=|\bar{\beta}_6|<\cdots\,,
\end{equation}
until the condition $\bar{\beta}<2$ (or equivalently $\beta<1$) is violated.\footnote{Note that when $\bar{\beta}<2$ ($\beta<1$) the Kasner transition process will be triggered.} We highlight that in \eqref{KasOsciAmp}, the symbol ``$<$'' corresponds to the Kasner inversion process, while ``$=$'' denotes the Kasner reflection process. Therefore, the processes of Kasner inversion and reflection occur alternately, resulting in the oscillation-amplification phenomenon observed in~\cite{Sword:2022oyg}.

\begin{figure}[H]
\centering
\includegraphics[width=0.75\textwidth]{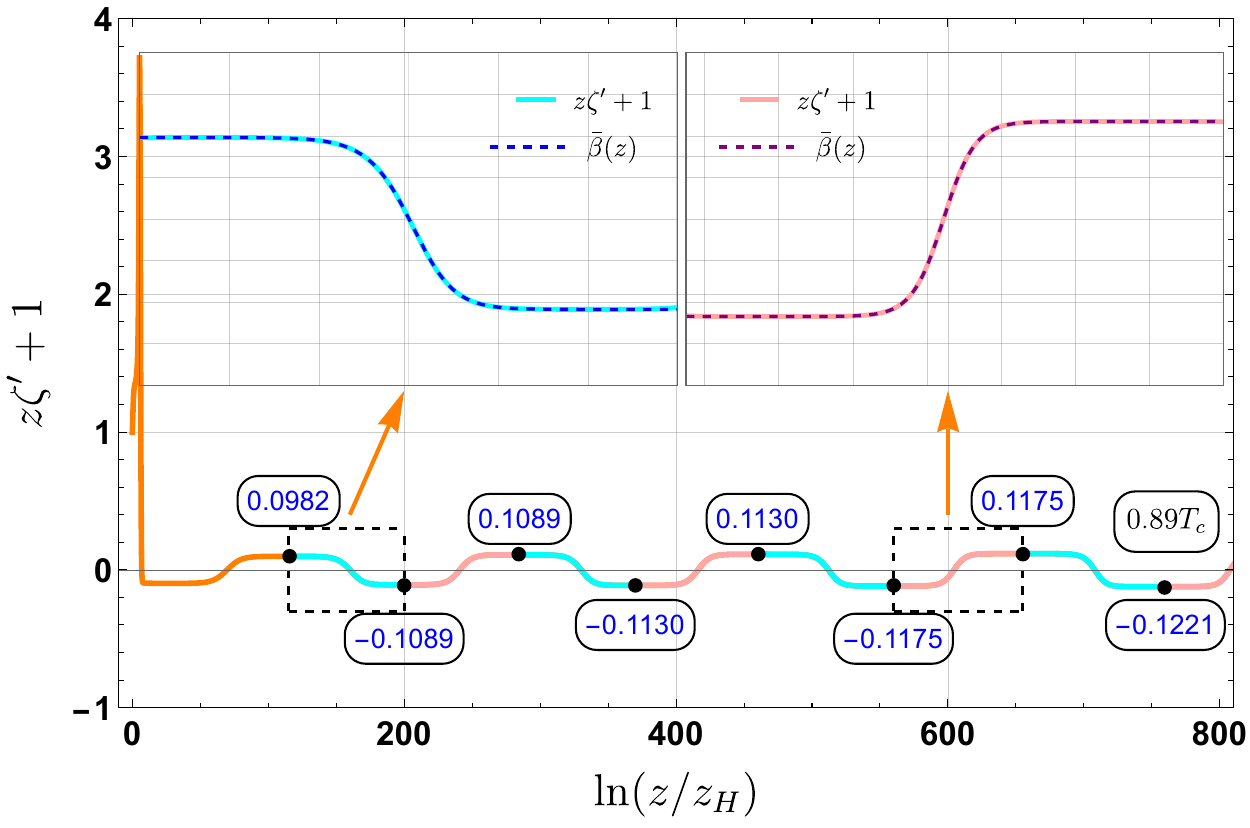}
\caption{The far interior evolution for $(z\zeta'+1)$ by solving the complete equations of motion~\eqref{EoM:zeta}-\eqref{EoM:h} at $T\approx 0.89T_c$ with $d=3$, $q=\sqrt{2}$ and $L=1$. The solid curve around horizontal axis demonstrates a large number of oscillations by virtue of the many plateaus corresponding to different Kasner epochs. The exponent $\bar{\beta}=\beta+1$ is shown explicitly in each Kasner epoch by fitting the numerical solution. Cyan color is for the Kasner inversion process and pink one for the Kasner reflection. \textbf{Insert}: We zoom in on the dynamics of the Kasner inversion and Kasner reflection processes, where the solid and dashed lines represent the numerical solution and analytic description, respectively.}\label{4d_beta_osci}
\end{figure}

A concrete example is depicted in Figure~\ref{4d_beta_osci}, 
from which numerous plateau oscillations are centered around $(z\zeta'+1)=0$, corresponding to value that sets $\beta=-1$. One can see the occurrence of Kasner inversion (cyan) and Kasner reflection (pink) alternately. We compare the profile of $\bar{\beta}=z\zeta'+1$ from the analytical solutions with the numerical ones of the full equations of motion (see Insert of Figure~\ref{4d_beta_osci}). The approximation of Kasner inversion~\eqref{InvAlphaSol} and Kasner reflection~\eqref{RefAlphaSol} is in excellent agreement to the numerical solution of the full equations of motion. 

Moreover, the alternation of all Kasner epochs are well predicted by~\eqref{KasLawOsci} and thus the oscillation amplitude grows towards the singularity at $z\rightarrow\infty$. As clear from Figure~\ref{4d_beta_osci}, the Kasner reflection process reflects the Kasner exponent with $\bar{\beta}_R=-\bar{\beta}$ and plays a key role for triggering such oscillation-amplification process. This is the main reason we call it Kasner reflection. Note that this intriguing behavior is independent of the spacetime dimension of the system, see~\eqref{KasLawOsci}. Additionally, according to our analytical laws~\eqref{KasLawAll}, this oscillation-amplification process would also occur near the fixed point $\beta=1$ in 4-dimensions ($d=3$) for which both the Kasner transition and Kasner inversion are described by almost the same dynamics, see~\eqref{TranDiffbeta1} and~\eqref{TranDiffbeta2} with $d=3$.

\section{Conclusion and Discussion}\label{Sec:DissAndCon}
We have studied the late time dynamics lying beyond the event horizon of the black holes with vector hair in asymptotically AdS spacetime. More precisely, we have considered the case with a massless charged vector field, for which the development of the vector hair necessarily removes any inner horizons. Due to the effects from vector condensate and U(1) gauge potential, it is characterized by a never-ending alternation of Kasner epochs towards the singularity, providing an explicit realization of Mixmaster chaos.

We have shown that the transitions between different Kasner epochs, and in particular the late interior time dynamics of bounces, can be captured analytically. More precisely, we have classified three types of Kasner alternations, including the Kasner transition, Kasner inversion and Kasner reflection, see Section~\ref{SubSec:KasStructure} for more details. Moreover, we have analytically derived the laws of the three Kasner transformations and verified the self-consistency of the analytic approximation. As shown in~\eqref{eqKas} and~\eqref{Kbeta}, each Kasner universe is determined by the parameter $\beta$. The transformation rule~\eqref{KasLawAll} covers all parameter space of $\beta$, and thus suggesting that we have clarified all possible kinds of Kasner transformation for the hairy black holes of~\eqref{GenAnsatz}. Our analytical results have been corroborated by numerical solutions to the full equations of motion. In Section~\ref{sec:Ncheck}, we have numerically solved the models for $d=3$ and $d=5$, which demonstrates that the dynamics of Kasner alternations is highly consistent with our analytical results~\eqref{KasLawAll}. 

Moreover, our analysis has provided a clear explanation of the dynamic behaviors found in the literature. As clear from~\eqref{KasLawPre}, the empirical equations summarized in~\cite{Cai:2021obq} are nothing but the Kasner transition and Kasner inversion in 4-dimensional spacetime ($d=3$). The dynamics with oscillation-amplification behavior observed in~\cite{Sword:2022oyg} was due to the combination of the Kasner inversion and Kasner reflection around the fixed point $\beta=-1$ (see \emph{e.g.} Figure~\ref{4d_beta_osci}). This intriguing behavior is independent of the spacetime dimension of the system, see~\eqref{KasLawOsci}. Moreover, one finds from~\eqref{TransitionOrderAnalysis3} that the Kasner reflection is triggered by the charged vector field ($q\neq 0$). One will generally not see such dynamics in the neutral vector case by setting $q=0$. Nevertheless, according to our analytical laws~\eqref{KasLawAll}, the oscillation-amplification process would also occur near $\beta=1$ in 4-dimensions ($d=3$) since the Kasner transition and Kasner inversion are described by almost the same dynamics when $d=3$ as shown in~\eqref{TranDiffbeta1} and~\eqref{TranDiffbeta2}.

In the present study, we have limited to the massless vector field. However, our analysis can be generalized to massive vector field. While the mass term was argued to be negligible in the far interior dynamics for neutral vector field~\cite{DeClerck:2023fax}, it appears that the mass term $m^2\rho^\mu\rho_\mu^{\dagger}$ becomes important once the exponent $n_{u}>2(d-1)$~\cite{Cai:2021obq}, or equivalently, $\frac{2\beta}{1+\beta}>d-1$, see~\eqref{NSol}. Therefore, when the exponent $\beta$ of a Kasner epoch falls into the range $-\frac{d-1}{d-3}<\beta<-1$, the contribution from the mass term could affect the Kasner dynamics, in particular, for the Kasner reflection.\,\footnote{Meanwhile, the mass term seems to play a key role using cosmological billiards formulated as a hyperbolic billiards problem~\cite{Henneaux:2022ijt}.} It will be interesting to understand the role of massive term and other general interactions in the interior dynamics. There have been some studies on the internal structure of black holes with rotation~\cite{Gao:2023rqc} and higher-derivative corrections~\cite{Bueno:2024fzg,Devecioglu:2023hmn}. Extending our approach to these cases would be interesting. 
Moreover, it will be desirable to understand the holographic meaning of the
black hole interior and to find a way to probe the interior dynamics~\cite{,Caceres:2023zft,Mansoori:2021wxf,Ceplak:2024bja,Arean:2024pzo}. Last but no least, all possible kinds of Kasner transformations can be collapsed into a single formula~\eqref{kastrad} in general spacetime dimensions. For $d=3$, it is precisely the case due the collision against a gravitational wall using the description in terms of cosmological
billiards. It is worth understanding~\eqref{kastrad} from cosmological
billiard for which there could be more than just the gravitational walls in higher dimensions. We hope to return to these issues in future work.

\section{Acknowledgements}
We thank Marc Henneaux for helpful discussions and comments. This work was supported by National Key Research and Development Program of China Grant No. 2020YFC2201501, and the National Natural Science Foundation of China Grants No. 12122513, No. 12075298, No. 11991052 and No. 12047503.

\bibliographystyle{JHEP}
\bibliography{refs} 

\end{document}